\documentclass[sigconf, usenames, dvipsnames]{acmart}

\AtBeginDocument{%
  }

\setcopyright{acmcopyright}
\copyrightyear{2018}
\acmYear{2018}
\acmDOI{XXXXXXX.XXXXXXX}

\acmConference[Conference acronym 'XX]{Make sure to enter the correct
  conference title from your rights confirmation emai}{June 03--05,
  2018}{Woodstock, NY}

\acmPrice{15.00}
\acmISBN{978-1-4503-XXXX-X/18/06}

\usepackage{float}
\usepackage{wrapfig}
\usepackage{xcolor}

\copyrightyear{2023}
\acmYear{2023}
\setcopyright{rightsretained}
\acmConference[CHI EA '23]{Extended Abstracts of the 2023 CHI Conference on Human Factors in Computing Systems}{April 23--28, 2023}{Hamburg, Germany}
\acmBooktitle{Extended Abstracts of the 2023 CHI Conference on Human Factors in Computing Systems (CHI EA '23), April 23--28, 2023, Hamburg, Germany}\acmDOI{10.1145/3544549.3585738}
\acmISBN{978-1-4503-9422-2/23/04}

\begin{document}

\acmBooktitle{}

\title{HoloTouch: Interacting with Mixed Reality Visualizations Through Smartphone Proxies}

\author{Neil Chulpongsatorn}
\affiliation{%
  \institution{University of Calgary}
  \city{Calgary}
  \country{Canada}
}
\email{thobthai.chulpongsat@ucalgary.ca}
\orcid{0000-0002-6283-7573}

\author{Wesley Willett}
\affiliation{%
  \institution{University of Calgary}
  \city{Calgary}
  \country{Canada}
}
\email{wesley.willett@ucalgary.ca}

\author{Ryo Suzuki}
\affiliation{%
  \institution{University of Calgary}
  \city{Calgary}
  \country{Canada}
}
\email{ryo.suzuki@ucalgary.ca}

\renewcommand{\shortauthors}{Chulpongsatorn, Willett, Suzuki}

\begin{teaserfigure}
\includegraphics[width=\textwidth]{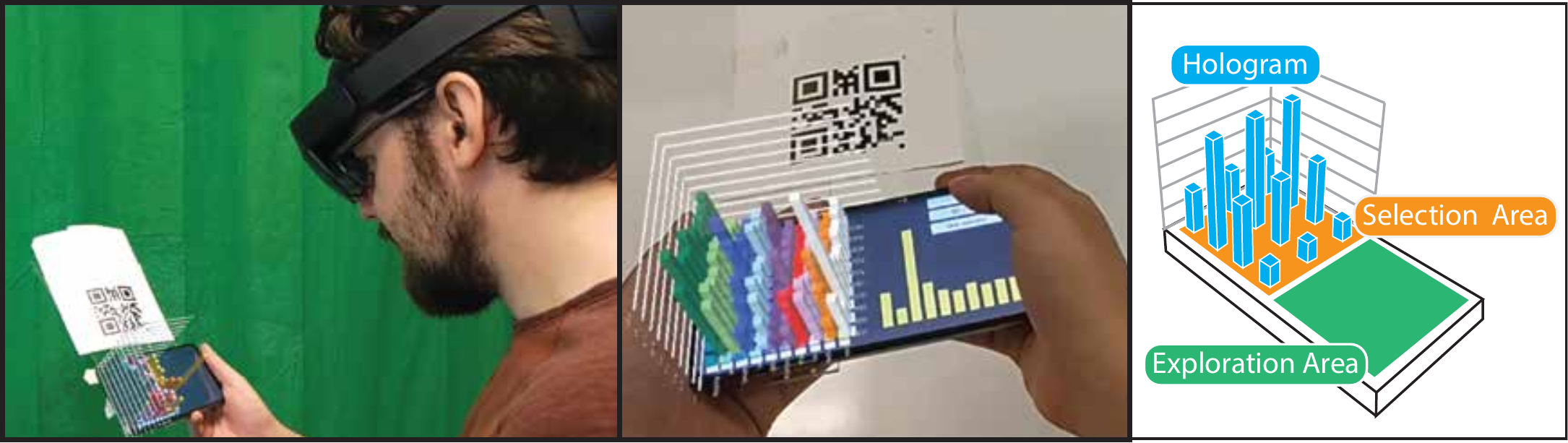}
\caption{The system is built in two parts: (1) The handheld smartphone proxy that handles user interaction, and (2) a holographic visualization that is rendered on top of the proxy. The user can create a selection by tapping the visual marks at its base (orange), then explore that subset of data in the green area using various features.}
\Description{Left: a person wearing hololens 2 holding a phone. There is a 3d holographic barchart that is mounted onto the pohone. Middle: Shows a clearer image of the phone with holographic bar chart. The hologram is mounted on the left of the phone. A 2D-onscreen barchart is renderered on the right side of the phone screen. Right: A figure illustrates how the system works. Blue barchart is drawn on top of a rectangular cube in the shape of a phone. The chart is labled "hologram". The area under the bar chart is highlighted in orange and labeled "selection area". the remaining area is highlighted in green and labeled "exploration area".   }
\end{teaserfigure}

\begin{abstract}
We contribute interaction techniques for augmenting mixed reality (MR) visualizations with smartphone proxies. By combining head-mounted displays (HMDs) with mobile touchscreens, we can augment low-resolution holographic 3D charts with precise touch input, haptics feedback, high-resolution 2D graphics, and physical manipulation. Our approach aims to complement both MR and physical visualizations. Most current MR visualizations suffer from unreliable tracking, low visual resolution, and imprecise input. Data physicalizations on the other hand, although allowing for natural physical manipulation, are limited in dynamic and interactive modification. We demonstrate how mobile devices such as smartphones or tablets can serve as physical proxies for MR data interactions, creating dynamic visualizations that support precise manipulation and rich input and output. We describe 6 interaction techniques that leverage the combined physicality, sensing, and output capabilities of HMDs and smartphones, and demonstrate those interactions via a prototype system. Based on an evaluation, we outline opportunities for combining the advantages of both MR and physical charts.
\end{abstract}



\maketitle

\section{Introduction}
Advances in mixed reality (MR) technologies have opened up exciting new opportunities for data visualization and analysis~\cite{ens2021grand}. 
Immersive data representations allow individuals to spatially explore and analyze complex data sets~\cite{hubenschmid2021stream} as well as directly embed data onto the real world~\cite{willett2016embedded}. 
However, existing MR visualization approaches~\cite{buschel2019augmented, bravo2021watch}, have key limitations in input, output, and interaction perspectives.
Interacting with current HMDs requires considerable effort that relies largely on air gestures, which can be imprecise due to relatively unreliable tracking technologies and the lack of tangible and tactile feedback.
This can also cause fatigue, inducing the ``gorilla arm'' effect~\cite{ens2021grand}.

The goal of this paper is to address these contemporary limitations of MR data visualizations by incorporating {\it smartphones as interactive tangible proxies} for interactions.
Our approach augments MR charts by leveraging mobile devices' precise touch input, audio and haptics feedback, high-resolution 2D graphics, and physical manipulation, all of which can complement the gestural interactions used for most current MR visualizations. The combination of smartphones and HMDs is not new and such interactions have been explored in the field of cross-device computing domains~\cite{zhu2020bishare}. 
However, despite papers showing the promise of such combined interactions~\cite{hubenschmid2021stream, langner2021marvis}, the space of possible interactions enabled by adding mobile devices to MR has seen little exploration.

In this paper, we contribute an initial design space exploration and qualitative evaluation that examines the potential for smartphones to serve as interactive tangible proxies for hand-scale mixed reality data visualizations.
Our main contribution includes: 
(1) A set of six example interaction techniques that leverage the combined physicality, sensing, and output capabilities of HMDs and smartphones.
(2) A demonstration of these techniques via an interactive prototype that combines smartphones and HMDs (specifically the Microsoft Hololens 2 \cite{hololens}), which examines how the presence of precise tangible input/output compares to purely HMD-based systems.
and (3), a preliminary evaluation where we outline opportunities for new immersive tools that combine the advantages of both MR and physical charts.

\section{Related Works}

\subsection{Mixed Reality Visualizations}
Augmented and mixed reality interfaces enable new possibilities for interactive data visualization and analysis~\cite{ens2021grand, willett2016embedded}. 
Existing approaches that leverage mixed reality visualizations can be largely categorized as (1) embedding data onto physical data referents to provide better association in the real world, such as situated~\cite{martins2021augmented, lee2012cityviewar} or embedded visualization~\cite{willett2016embedded}, and (2) 3D immersive visualization that leverages spatial manipulation and exploration in physical spaces~\cite{cordeil2019iatk, sicat2018dxr}. 
The first category focuses on the relationship between physical data referents and visualizations~\cite{bach2017drawing}.
For example, existing work on situated and embedded visualizations typically augment static real world visualizations by overlaying data in AR/MR~\cite{chen2020augmenting}, or help users to associate visualizations with geo-spatial data referents~\cite{martins2021augmented, suzuki2020realitysketch, lee2012cityviewar, bach2017drawing}.

The second category focuses more on the exploration of new interaction techniques enabled by spatial-awareness, which is more relevant to our focus.
AR/VR visualizations have many advantages, such as enabling natural collaboration for data analysis~\cite{butscher2018clusters, donalek2014immersive}, enabling flexible manipulation for complex multivariate data~\cite{cordeil2017imaxes}, and facilitating spatial navigation and exploration of geo-temporal data~\cite{ssin2019geogate, merino2017impact, coulter2007effect}.
Recent work has also examined the effectiveness of  3D immersive visualization, compared to traditional 2D visualization~\cite{kraus2020assessing, yang2018origin}.
However, immersive visualizations, especially ones using mixed reality HMDs, still suffer from many limitations and challenges.
One of the limitations is the lack of tactile feedback and physical manipulation capability, which makes it difficult for viewers to interact with data~\cite{danyluk2020touch}.

\subsection{Cross-Device Interaction for Mixed Reality}
Previous works have extensively explored cross-device computing~\cite{  grubert2015multifi,  budhiraja2013interaction,   arora2018symbiosissketch} as a way to address the limitation of HMD's.
In many of these examples, mobile devices expands and enriches the design space of interactions with HMDs. For example, we could leverage precise touch input of these devices to manipulate virtual objects \cite{knierim2021smartphone, dorta2016hyve, mohr2019trackcap, surale2019tabletinvr}, complement low-resolution 3D graphics with high-resolution screens \cite{serrano2014identifying}, and support rich tangible and spatial manipulations~\cite{darbar2021exploring, normand2018enlarging, dorta2016hyve, millette2016dualcad}.
Zhu and Grossman's BISHARE~\cite{zhu2020bishare} synthesized many of these techniques and categorized them into phone-centric and HMD-centric interactions for 3D virtual object manipulation.

Inspired by these researches, the visualization community also recently started exploring combining HMD and mobile devices for immersive data analysis.
For example, Hubenschmidt et al.'s STREAM applies a combination of HMD and a tablet to enable an embodied interaction with physically distributed MR multi-view visualizations~\cite{hubenschmid2021stream}. Huang et al. illustrated a framework for AR-smartphone visual analytics \cite{huang2022sparvis}, but their explorations are largely scoped within 2D holographic visualizations. 
Bushen el al. investigated the use of smartphone as pan and zoom interaction proxy for 3D visualizations \cite{buschel2019investigating}, however a characterization of the larger interaction space remains to be seen.  
Most relevant to our work, Langner et al.'s MARVIS, introduced a set of visual augmentation techniques for mobile data visualization~\cite{langner2021marvis}.
These techniques include seamless extensions of the screen content, superimposed 3D visualizations, and off-screen content for supplemental information such as legend and menus.
Their work focused largely on demonstrating these \textit{visualization techniques}. An exploration of \textit{interaction techniques} and in-depth insights about the augmentation through user evaluations is still missing. 
Our work contributes to the design and evaluation of the interaction space around combined HMD-smartphone MR visualizations.
In particular, we are interested in how the set of features enabled by smartphones, such as precise input, physical spatial manipulation, and audio/tactile feedback, could support data reading and exploration using handheld 3D holographic visualizations.

\subsection{Tangible Interactions with Data}
Our work takes inspiration from tangible interactions for data visualizations~\cite{cordeil2020embodied}.
In particular, Jansen et al. examined the use of physically fabricated, assembled, or actuated representations of data as alternatives to screen or AR/VR-based data visualizations~\cite{jansen2015opportunities}. 
Data physicalizations can outperform their on-screen counterparts in three ways: touch, manipulation, and realism \cite{jansen2013evaluating}. 
However, the benefits of being able to physically touch and manipulate data representations are still usually outweighed by the dynamic nature of virtual representations, which can be easily updated in response to user inputs through programmable interactions \cite{danyluk2020touch}. 
Some recent works have explored the combination of tangible interactions and AR to support embodied interactions~\cite{suzuki2020realitysketch, suzuki2019shapebots, monteiro2023teachable}, while allowing rich visual feedback with AR displays.
Our work also seeks to strike a balance between these approaches. We examine how physical proxies might pair some of the tangible benefits of data physicalizations, while still supporting dynamic interactions via MR for small handheld visualizations. 
We hypothesizes that this combination of tactile manipulation and touch input, anchored with holographic visualizations, could provide similar benefits to data physicalizations.

\section{Design Justification}

\begin{figure}[ht]
\includegraphics[width=\linewidth]{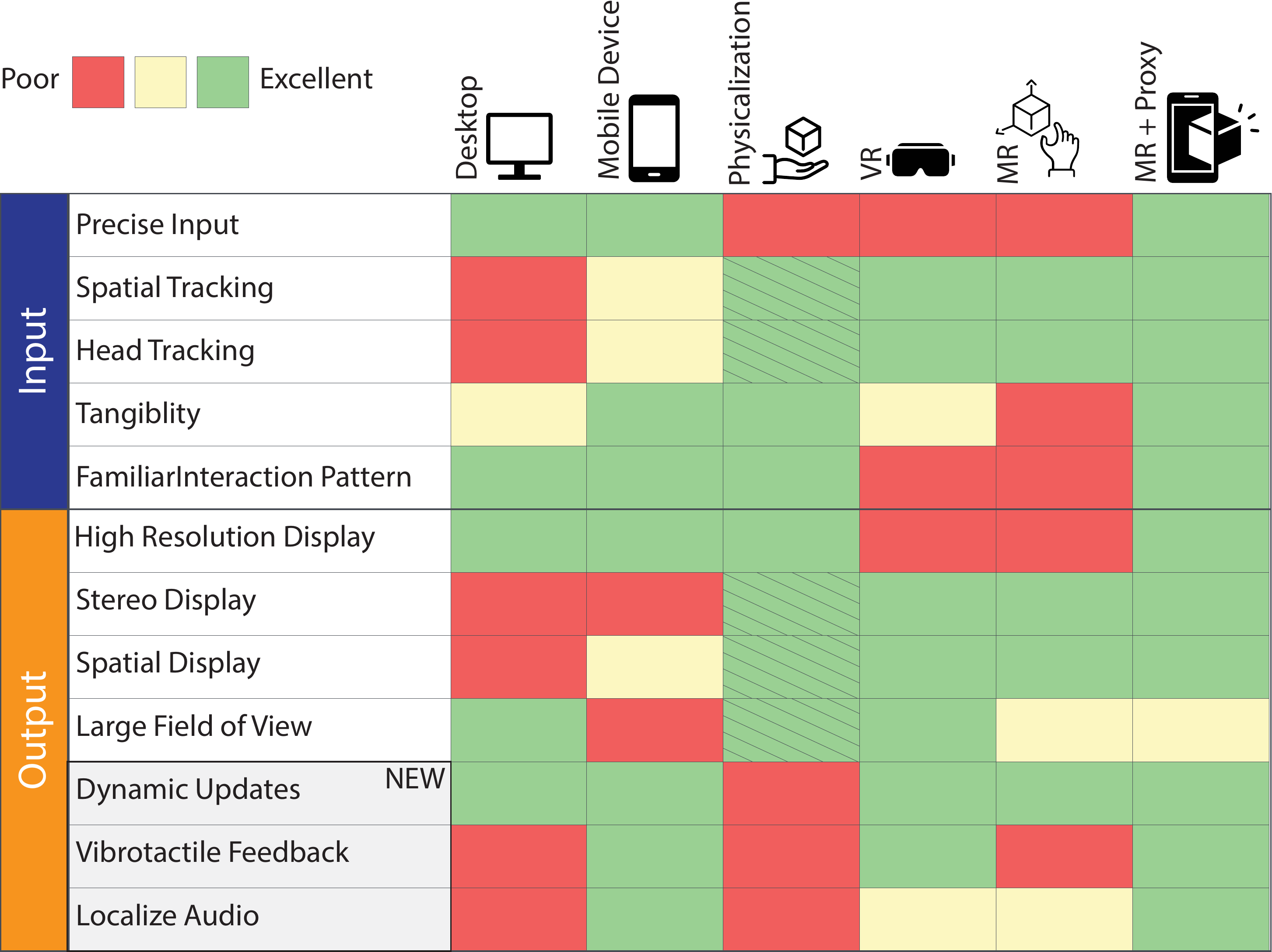}
\centering
\caption{An overview of the capabilities of various visualization platforms including desktop, mobile, physicalization, VR, MR HMD, and MR HMD + device variations(*an expansion of Zhu and Grossman's original comparison of Mobile Devices and HMDs~\cite{zhu2020bishare}}   
\Description{A table is shown. From top to bottom, the row is labeled: "precise input", spatial tracking", "head tracking", "tangibility", "familiar interaction pattern", "high resolution display", "stereo display", "spatial display", "large field of view", "nynamic updates (new)", "vibrotactile feedback", "localized audio". From left to right, the rows are labled as: "desktop", "mobile device", "physicalization", "VR", "MR", "MR + proxy".}
\end{figure}

We extend Zhu and Grossman's analysis of HMDs and smartphones~\cite{zhu2020bishare} to characterize the input and output capabilities of various common visualization modalities. 
Input capabilities include (1) precision and accuracy of the input, (2) spatial tracking of hand and body, (3) head tracking, (4) tangible and embodied manipulation, and (5) familiarity of the interactions. 
Meanwhile, output capabilities include the presence of (1) a high-resolution display, (2) support for stereoscopic depth, (3) a spatially-aware content display, (4) a large field of view, (5) the ability to dynamically update, (6) vibrotactile feedback, and (7) localized audio feedback. 
In the following section, we briefly discuss and compare each modality. 

\subsubsection*{\textbf{Desktop.}} Visualizations on 2D monitors can render excellent high-resolution details. However, they are inherently limited by their incompatibility with spatial interactions and 3D inputs and outputs without a specialized equipment. 

\subsubsection*{\textbf{Mobile devices.}} Similar to desktop screens, mobile devices also support high-resolution 2D graphics. On top of that, the devices are equipped with rich audio output, vibrotactile feedback, and an assortment of sensors and mechanism depending on their model. They can also partially track their own position using internal sensors such as cameras or accelerometers, although the precision of spatial tracking and understanding is still limited. 

\subsubsection*{\textbf{Physicalization.}} Tangible data representations can enable rich embodied and tactile data exploration, which are difficult to approximate with on-screen devices.
Most current physicalizations are limited by the static and inflexible nature of the physical objects themselves, though some work have examined dynamic physicalization through mechanical actuation \cite{taher2016investigating}.
However, the limited generalizability and flexibility, as well as the cost of producing them, likely limits physicalizations' utility for exploratory or analytical data exploration~\cite{jansen2015opportunities}.

\subsubsection*{\textbf{Virtual reality.}} Immersive data visualizations in virtual reality (VR) allows for spatial navigation and exploration of 3D visualizations, providing a much clearer sense of space and scale~\cite{hayatpur2020datahop,liu2020design}. Although recent advances in display technologies have dramatically increased the resolution of VR displays, they are still outperformed by 2D desktop displays and mobile devices~\cite{kraus2020assessing}. Additionally, interactions with VR data representation typically relies on handheld controllers. More recently, hand tracking have also been built into headsets such as the Meta Quest\cite{oculus}. Although it still remains relatively imprecise, and it can be difficult for viewers to interact with detailed virtual objects and spaces~\cite{danyluk2020touch, arora2017experimental}.

\subsubsection*{\textbf{HMD mixed reality.}} Another immersive visualization approach relies on head-mounted mixed reality displays, which render 3D visualizations onto existing physical environments. Mixed reality approaches promise many of the benefits of VR visualization, in addition to the ability to view data in the context of real-world settings and tasks~\cite{white2009sitelens, prouzeau2020corsican}. 
However, MR HMDs suffer from numerous limitations due to their technical immaturity; these include narrow field-of-view, poor object tracking, and occlusion management. Moreover, most current MR displays rely on hand-tracking and air gestures for input, resulting in less input precision than controller-based VR solutions.  This altogether removes the possibility of enhancing user interactions with vibrotactile~\cite{prouzeau2019scaptics} and audio feedback through controllers.

\subsubsection*{\textbf{HMD mixed reality + mobile devices.}} While none of these approaches are superior in all respects, mobile devices and mixed-reality HMDs have complementary characteristics. As a media for visualization, mobile devices are limited to their flat-2D screen space. However, phones and tablets offer not only tangibility, but also precise and high-resolution 2D input and output, which makes them great as input devices. On the other hand, HMDs provide larger spatially-registered 3D imagery with stereo depth cues that mobile devices lack. 
The set of possible interaction techniques using this combination is still relatively unexplored in a visualization context. 


\section{MR Visualization Interactions using Smartphone Proxies} \label{proxy}
In this section, we explore the use of mobile devices as interaction proxies for MR visualizations. We discuss the space that our designs are based upon and explore six proxy interactions with 3D data visualization (Figure \ref{fig:interactions}): discrete select, axis select, physical manipulation, vibrotactile feedback, summarization, and projection.

\subsection{Interaction Design Space}
Introducing a mobile device as a proxy for physical interactions enables a variety of new opportunities for interacting with MR data visualizations. We first review a set of common visualization interactions, then describe six specific interaction techniques that leverage the combined affordances of MR and smartphones. 
We base our work on six of the general categories (\textit{select, explore, encode, abstract/elaborate,} and \textit{filter}) inspired by Yi et al~\cite{yi2007toward}. 

For consistency and simplicity, we focus our discussion around 2.5D bar charts as they represent one of the most straightforward and useful 3D chart idioms. Such charts, which typically visualize one quantitative value across two categorical ones, have been featured extensively in prior work on physicalization and immersive analytics ~\cite{jansen2013evaluating, danyluk2020touch}. 
They have also served as the basis for dynamic physicalization prototypes ~\cite{taher2016investigating}, VR tools ~\cite{liu2020design}, and AR data visualizations ~\cite{langner2021marvis}. The 2.5D nature of these charts also means that each mark has a unique 2D position in X and Y---a configuration that maps particularly well to 2D interactions on a smartphone screen. 

\begingroup 
\setlength{\columnsep}{12pt}

\begin{figure*}[ht]
\centering
\includegraphics[width=1\textwidth]{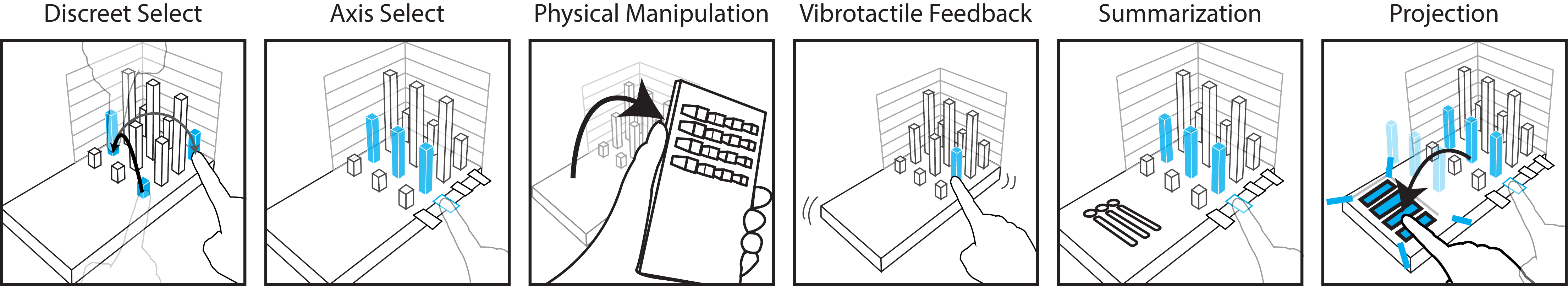}
\caption{ An overview of smartphone proxy interactions with holographic visualization}
\Description{6 pictures are shown, and includes a phone and a holographic barchart on top of it.. 1: discreet select; fingers points onto individual highlighted bars. 2: axis select, figer points at axis label, a row is highlight. 3: physical Manipulation: a hand is show to move the phone, and the visualization moves with it. 4:vibrotactile feed back. A finger points on to a bar, vibrations are drawn around the phone. 5: summarization: a finger taps on a row and a summary is shown on the side of the phone without the hologram. 6: a figer drags a row of 3D bars onto the phone, where it turns into 2D bars.   }
\label{fig:interactions}
\end{figure*}

\subsubsection*{\textbf{Select}} 
Interactions typically highlight marks or subsets of the visualization, allowing viewers to keep track of specific data marks and often enabling further interactions (filtering, sorting, copying, etc.) with the selected set. 
In most desktop and mobile visualization tools, selections rely either on direct pointing or box/brushing interactions, all of which can be challenging in MR. Using a smartphone screen can simplify these interactions by providing a stable touch target with precise input tracking. We consider three different possible selection interactions that leverage the proxy.
Viewers can use \textbf{discrete selection} and interact with individual marks by directly tapping the smartphone screen beneath the MR visualization. Using the smartphone screen rather than air gestures provides far greater precision with detecting these selections. 
The smartphone screen can also support interactions that use other visualization elements including axes and legends. Viewers can use \textbf{axis selection} by taping or sliding a finger on the axis of a chart to create selections across a dimension in the visualization.

\subsubsection*{\textbf{Explore}} Physical and MR representations of data make it possible to use head movement, physical locomotion, and active spatial manipulation to explore the visualization. Adding a physical proxy has the potential to make some of these interactions (particularly ones that involve manual re-positioning of the visualization) easier by providing a real-world handle that actively adjusts the virtual chart. 
The viewer can move and rotate the visualization through \textbf{physical manipulation} with the proxy. This allows viewers to re-position the charts relative to their head and hands in much the same way as a physical bar chart. 

\subsubsection*{\textbf{Encode}} The addition of a physical proxy also introduces new opportunities for data encodings that leverage interaction. 
Visualizations can double-encode data values to aid perception and interaction by using the additional output modalities of the proxy. 
Vibration motors in smartphones makes it possible to use \textbf{vibrotactile feedback} patterns or intensity to encode quantitative data.
The vibration can be used to encode a measurement of a data mark, or even the difference between two data mark.

\subsubsection*{\textbf{Abstract/Elaborate}} Many visualizations provide interactions for controlling the level-of-detail. The addition of high-resolution smartphone displays to mixed-reality visualizations 

\noindent creates opportunities to present these kinds of abstracted or elaborated information in ways that increase their legibility.
The \textbf{summarization} interaction. Allows the viewers to use smartphone-mediated interactions with a holographic chart to display summary statistics and aggregates based on a selection. 

\subsubsection*{\textbf{Filter.}} Filter interactions allow viewers to hide slices of larger datasets, leaving only a subset visible.
A single series of visual marks from a larger holographic bar chart can be \textbf{projected} onto the high resolution smartphone display to make it possible to examine that data by itself, while also retaining the context of the original chart. This can help with viewing pieces of the visualization that are occluded in 3D form.

\endgroup 

\subsection{Proof-of-Concept Prototype}
We implemented these interactions using a Microsoft Hololens 2 and an Android smartphone (One Plus 7t Pro~\cite{oneplus7tpro}) as the physical proxy.
Both devices run a custom visualization application made with Unity~\cite{unity} and synchronized via socket.io~\cite{socketio} and Node.js~\cite{nodejs} server on a local network. 
The mobile client handles touch input, provides vibrotactile feedback, and renders 2D visualizations, text, and other information on its display.
The mobile client sends any interaction it detects from the user to the Hololens client that will update the 3D visualization accordingly.
The Hololens uses Microsoft's QR code NuGet package~\cite{nugetqr} to track the phone's position.

\section{Preliminary User Evaluation}

\begin{wrapfigure}{r}{2.8cm}
\vspace{-.4cm}
\includegraphics[width=2.8cm]{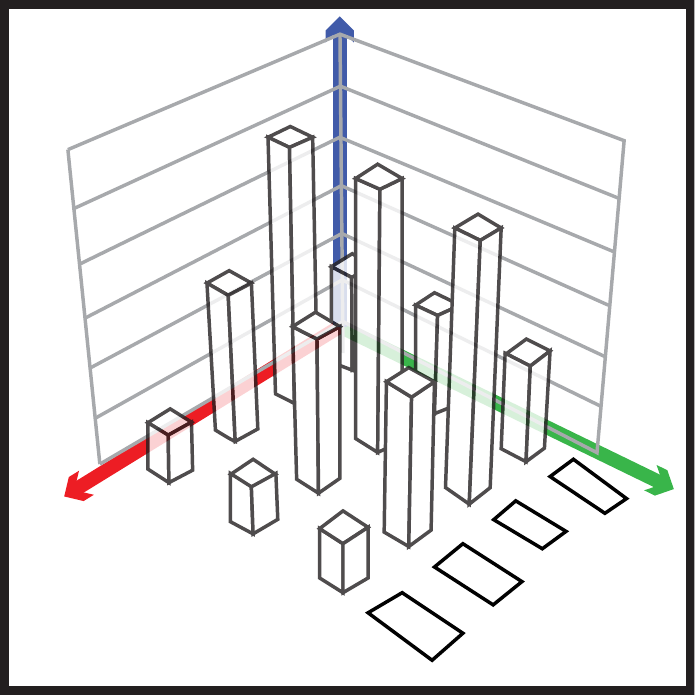}
\Description{A 3d barchart. A blue arrow points upward. A red arrow points to the bottom left. A green arrow point to the bottom right.}
\end{wrapfigure}


Our study takes inspirations from existing papers that compare the usability of different visualization modalities~\cite{jansen2013evaluating, danyluk2020touch}. The goal of this study is to deploy the prototype as a mean to explore the design space. We are interested to see what kind of interactions are promising in this setup. We also what to know how the smartphone proxy can assist with data visualization tasks, as well as their benefits and limitations. 
We chose three data sets that contains these three dimension: location, year, and a quantitative dimension (includes car mortality rate, carbon dioxide emissions, and military expenditure as a percentage of GDP). These are a subset of common data that were used in previous evaluations of data physicalization approaches ~\cite{jansen2013evaluating, danyluk2020touch}. These visualizations are recreated in mixed reality. All of the visualizations encode year on the horizontal axis (green), countries on the depth axis (red), and the numerical measure on the vertical axis (blue). We implemented two as MR  with a physical proxy charts to give user more time with the system, and one as a pure MR chart without the proxy as a baseline. 

Similar to prior work \cite{jansen2013evaluating, danyluk2020touch}, we asked participants to perform the same three data analysis tasks on the visualizations. These covered an array of basic operations in data analysis, which includes: (1) a \textbf{range} task to identify the local minima and local maxima within a certain dimension; (2) an \textbf{order} task to sort the values within a dimension in ascending order; and (3) a \textbf{compare} task to find three distinct country-year pair and determine which has the lowest value.

\begin{figure}[ht]
\includegraphics[width=\linewidth]{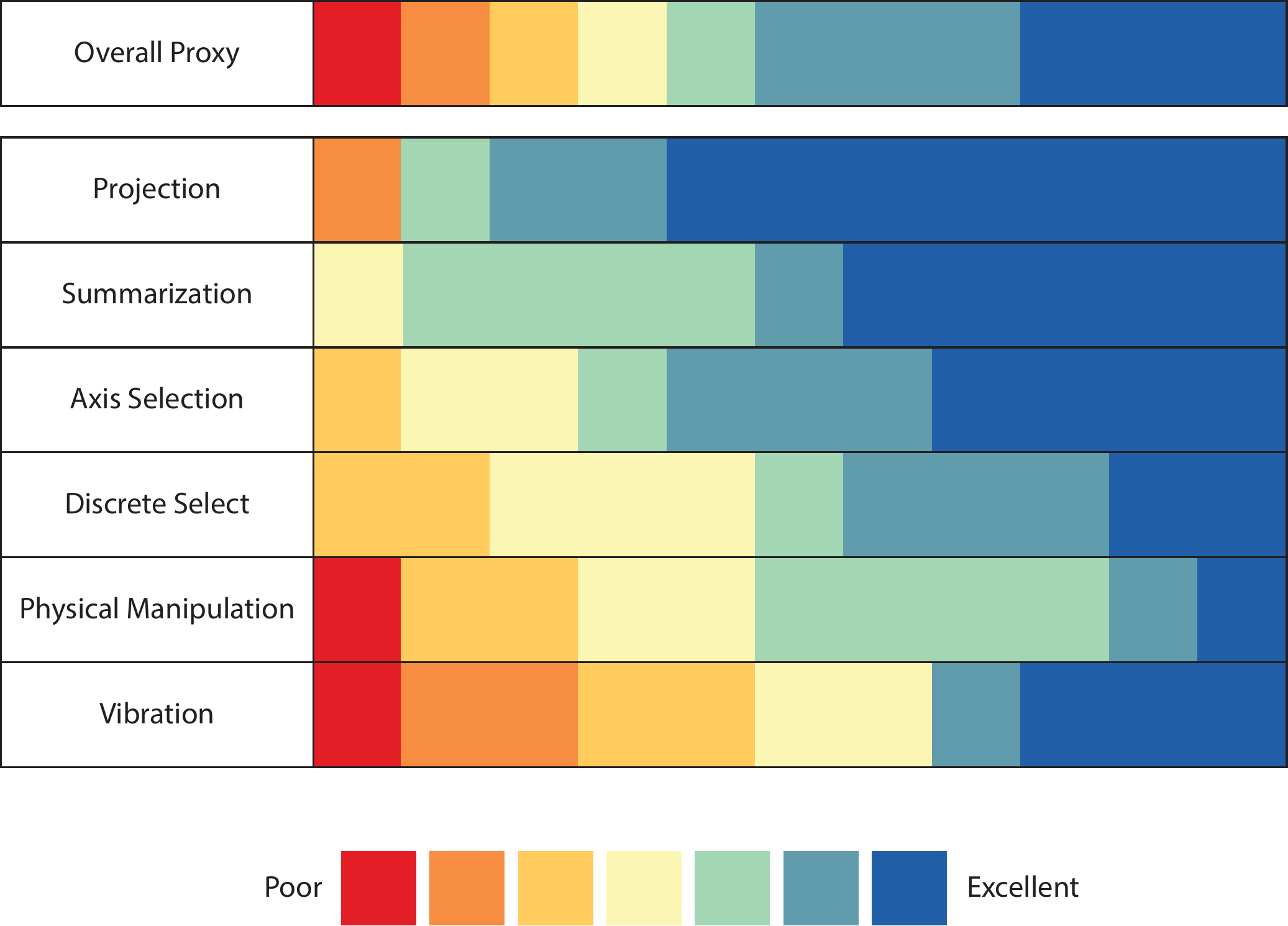}
\centering 
\caption{Summary of user satisfaction survey with the overall system and each of the features.}
\Description{A chart is shown, colors are used to encode particiapant feedback on a likert scale. From top to bottom, the rows are labled "overall proxy", projection, summarization, axis selection, discrete select, physical manipulation, vibration. }

\end{figure}

\begin{figure*}[ht]
\centering
\includegraphics[width=1\textwidth]{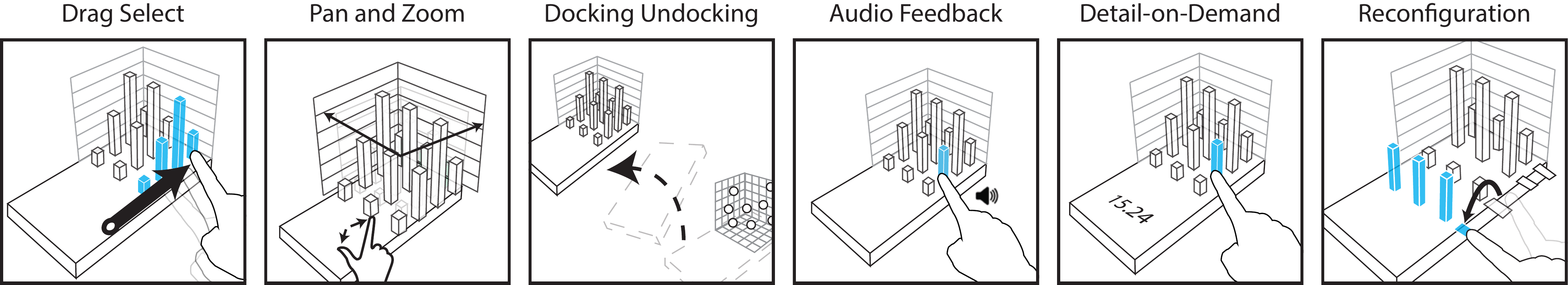}
\caption{Possible interactions to be explored in future work.}
\Description{6 images are illustrated, each includes a hand, and a phone with a 3D bar chart on top of it on one side. 1: drag select. A finger is drawn to drag across the phone. The bars it passes are highlighted. 2: Pan and zoom. A hand points with thumb and index finger perpendicular to each other. An arrow is drawn pointing to the tip of each of the finger. The 3D barchart is drawn as larger than the phone, with arrows point in the same direction as before. 3: docking unlocking. two 3D visualizations are drawn. A phone is shown to move between them. 4: audio feedback. a finger points at a bar and and audio icon is drawn beside it. 5:detail on demand. a figure points at a bar and a number appears on the empty side of the phone. 6: reconfiguration: a finger is draw to drag a row along the phone.   }
\label{fig:future}
\end{figure*}

\subsection{Participants and Procedure} We recruited eleven adult volunteers (P1 - P11) through word of mouth and snowball sampling. Two participants were color blind (P8 and P11); the remaining participants had normal or corrected to normal vision. 
Five participants had extensive experience with VR, and six had little to none. 
One participant works with visualization at a professional level, nine were as students, and one had no experience. 
We calibrated the Hololens before each session, then presented the participant with the aforementioned 3D-bar charts --- one at a time. We quickly demonstrated each feature and gave them time to get used to the interactions before the experiment. We used a different barchart for each trial to ensure that participants' performances would not be affected by their familiarity with a data set.
For each trial, participants performed  range, order, and compare tasks and said their answer out loud while the researcher inputs their responses on an answer sheet. After the experiment, we gave them a questionnaire asking them to rank their satisfaction for each of the features on a seven-point Likert scale and provide a rationale for their evaluation.


\subsection{Results} The participants' feedback for the implementation were generally positive (4.9). Projection was a favourite with only one negative response and an average user satisfaction score of 6.2. This is likely because this interaction allowed the participants to easily and clearly view an occluded row of the chart. Following that in descending order:  summarization (5.9), axis selection (5.6), context menu (5.5), discrete selection (5), physical manipulation (4.4), and finally vibration (4.2). 
Four participants commented on how it helped with their exploration (P1, P4, P5, P8). P1 stated that they preferred having something tangible in their hand to move the visualization around. P3 and P6 mentioned how they liked the precision that the proxy provided. Specifically, they were able to navigate the on-screen menu quickly and comfortably. P3 talked about how the proxy allowed them to easily get a precise number, but also said that it felt like they did not need the 3D visualization at all and that they \textit{``really only need the labels and the ability to select the bars''}. Some participants experienced minor difficulties with misalignment and blurriness of the system, but were still generally positive about the smartphone proxy interaction overall (P7, P9). P10 and P11 explicitly stated that they preferred to explore the data visualization without the proxy. P10 had difficulty with the interactions and could not register their input consistently. Particularly, they mention how it felt like the hologram was visually blocking the phone, and was generally in the way of interacting with the proxy system. This is likely because the hologram can visually occlude the button depending on the viewing angle. P11 also remarked on how the hologram felt tacked on to the phone and preferred just the static visualization as a whole.

\section{Discussion and Future Works} \label{discussion}
Our initial study, along with our own experiences building and testing our prototypes highlight several remaining issues with current hardware that limits the approach, while also highlighting opportunities to further extend the concept.


\subsubsection*{\textbf{Visual Clarity}} Contemporary MR headsets like the Hololens 2 still lacks the resolution and pixel density of other types of digital displays. 
This means that holographic visualizations rendered via these systems cannot yet achieve high visual data densities, nor can it include detailed labels, axes, and other visual elements except at a coarse level. Despite our focus on relatively simple visualization designs, viewers using our prototype often found the text difficult to read. The translucent nature of the holograms also contributes to legibility issues. While resolution, contrast, and image clarity will undoubtedly improve in future iterations, the short-term solution is to increase the size of the visualizations. However, doing so either requires additional interactions to support large visualizations using a much smaller proxy. Larger proxies may also be considered, but its size and weight would likely introduce new interaction challenges. 

\subsubsection*{\textbf{Visualization-Specific Proxies}} We leveraged off-the-shelf smartphones as our physical proxies for our implementation. While these devices are ubiquitous and feature-rich, their design still introduces some limitations in sizes and aspect ratios, and almost all of their input and output are on the front of the device.
However, dedicated visualization-specific proxies could likely support more useful outputs. For example, dedicated proxies could be produced in sizes and shapes that are easier to use in multiple orientations. This can provide more ergonomic grips, touch points, and input sensors designed to support common visualization interactions. Side- and back-of-device touch support~\cite{baudisch2009back}, for example, could facilitate easier interactions with axes or allow viewers to interact with holograms without occluding data points. Similarly, edge-of-device display areas could make it possible to augment low-resolution holograms with higher resolution axes and labels. Meanwhile, richer and more powerful  vibrotactile motors and speakers distributed across the device could support stronger and more nuanced non-visual data encoding. Support for magnetic or interlocking couplings between proxies (as in Marquardt et al.'s SurfaceConstellations~\cite{marquardt2018surfaceconstellations}) could also make it possible to combine devices to enable visualizations at multiple sizes or to enable specific kinds of cross-device data transfer and interactions.

\subsubsection*{\textbf{Beyond Bar Charts}} Our explorations focused largely on discrete ``2.5D'' visualizations like 3D bar charts, where each visual mark corresponds unambiguously to a specific input area on the device. These kinds of representations are common both in recent immersive analytics and physicalization research. This scope also helps us focus more on this preliminary exploration of interactions within the space. However, we are aware that this scope is limited and that there are many different representations beyond just bar charts. We are excited to see exploration in interactions with vastly different visualizations such as continuous full-3D visualizations like 3D scatter plots or space-time cubes. Visualizations where individual input locations on a 2D screen may not map unambiguously to a single mark in the visualization will add interesting dimensions to our space and represent a promising area for future explorations. 

The lightweight nature of the physical proxy could be used to help enable interactions with full 3D visualizations in a variety of ways. For example, translation interactions could allow the phone screen to slide up through the visualization volume like a cutting plane (perhaps in response to a finger on the side of the device) allowing viewers to interact with points at any depth in the volume. Docking/undocking interactions could also be used to lock a visualization in space, allowing the proxy to be used as a cutting plane and interaction surface at a variety of angles. Vibrotactile feedback could also prove valuable here, helping viewers halt the cutting plane at depths with greater data density. 

Touch-and-slide interactions, like those used in many mobile maps applications could also be used to select items at varying distances above the device's screen. Unlike with 2.5D visualizations, viewers may wish to rotate or pivot full 3D visualizations with respect to the proxy, allowing them to see the data from different angles and enabling easier selections along the axes currently aligned to the display. More complex 3D visualizations such as network graphs and surface models likely present even more design opportunities.


\subsubsection*{\textbf{Design Space Expansion}}
As this paper is an initial exploration to interaction techniques with mobile proxy, there are still much left to be done. First sorting interactions, as listed by Yi et al. \cite{yi2007toward} as a common technique, is still unexplored. This is an important next step, especially since it would help with the occlusion problem that we observed in our study. We can also explore more proxy interaction techniques for each of the visualization interaction. For one we discussed only discrete selection techniques, but there are still continuous methods like brushing and linking with the proxy to be seen. There may also be more encoding methods that utilize other outputs from the phone such as sound and light. Lastly, we focused around a single visualization and single proxy in this paper, but we are interested to see how multi-device and multi-visualization as demonstrated by MARVIS \cite{langner2021marvis} may affect the interaction design space we proposed. We summarize some ideas that were inspired from our study in Figure \ref{fig:future} above. 


\section{Conclusion}
This initial exploration in using smartphone proxy already showcases some of the unique benefits of this hardware combination and highlights the breadth of interactions it might support. The space of possible visualization and proxy designs to support them is vast and holds numerous opportunities for future research. 
Combining MR and physical proxies also has the potential to support a much wider variety of visualization designs, including dynamic representations as well as situated and embedded ones. Combined with advances in mobile data processing and analysis, they have the potential to unlock a future in which rich, embodied interactions with data are possible almost anywhere.

\begin{acks}
This work was supported in part by the Natural Sciences and Engineering Research Council of Canada (NSERC) [RGPIN-2021-02857] and the Canada Research Chairs Program.
\end{acks}

\balance
\bibliographystyle{ACM-Reference-Format}
\bibliography{references}

\end{document}